# Enhanced Breakdown Voltage in β-$Ga_2O_3$ Schottky Diodes via Fast Neutron Irradiation and Electro-thermal Annealing

Saleh Ahmed Khan[1, a)], Sudipto Saha[2, a)], Ahmed Ibreljic[1], Stephen Margiotta[1], Jiawei Liu[2], Walid Amir[2] and Surajit Chakraborty[2], Uttam Singisetti[2], A F M Anhar Uddin Bhuiyan[1, b)]

[1]Department of Electrical and Computer Engineering, University of Massachusetts Lowell, Lowell, Massachusetts 01854, USA

[2]Department of Electrical Engineering, University at Buffalo, Buffalo, New York 14260, USA

[a)]S. A. Khan and S. Saha contributed equally to this work

[b)]Corresponding author email: anhar_bhuiyan@uml.edu

## Abstract

This study demonstrates a substantial enhancement of breakdown voltage in β-$Ga_2O_3$ Schottky barrier diodes (SBDs) through an approach that combines fast neutron irradiation with controlled post-irradiation electro-thermal annealing. Devices irradiated with 1 MeV neutrons at a high fluence of $1\times10^{15}$ n/cm² initially exhibited substantial degradation, including a drastic reduction in on-current and an increase in on-resistance. Electro-thermal testing, conducted through simultaneous current-voltage measurements while heating the devices up to 250 °C, resulted in significant recovery. After four cycles of electro-thermal testing, the devices demonstrated significant improvements in performance, with a substantial recovery of on-current and a reduction in on-resistance compared to the post-radiation condition, approaching pre-radiation levels. Most recovery occurred during the first two cycles, with diminishing improvements thereafter, indicating that thermally responsive radiation-induced traps were largely mitigated early in the process. Capacitance-voltage measurements revealed a substantial reduction in net carrier concentration, decreasing from $3.2\times10^{16}$ cm$^{-3}$ pre-radiation to $5.5\times10^{15}$ cm$^{-3}$ after the first electro-thermal testing cycle, indicating an over 82% reduction. Following the third cycle, the carrier concentration partially recovered to $9.9\times10^{15}$ cm$^{-3}$, reflecting a carrier removal rate of ~22 cm$^{-1}$.

The breakdown voltage exhibited a remarkable enhancement, increasing from approximately 300 V to 1.28 kV (a ~325% improvement) after the first electro-thermal testing, can be attributed to the reduction in net carrier concentration by compensating radiation-induced traps. Subsequent testing reduced $V_{br}$ slightly to 940 V due to partial recovery of carrier concentration, but it remained significantly higher than pre-radiation levels. These findings demonstrate the potential of combining neutron irradiation with electro-thermal annealing to significantly enhance the voltage blocking capability of β-$Ga_2O_3$ power devices, making them strong candidates for high-power applications in radiation-intense environments.

β-$Ga_2O_3$ has emerged as a promising ultrawide bandgap semiconductor for high-voltage electronic applications, owing to its large bandgap (~4.8 eV), high electric field (6-8 MV/cm), controllable n-type doping, and availability of large-area, cost-effective native substrates[1]. The intrinsic material robustness of β-$Ga_2O_3$, supported by the high atomic displacement energies of Ga and O atoms[2], suggests a strong resilience against displacement damage, positioning β-$Ga_2O_3$ as a promising candidate for high-power electronics operating in radiation-intense environments such as space, nuclear systems, and high-altitude avionics[3-6]. Among radiation sources, neutrons are especially damaging due to their high penetration depth and ability to displace atoms in the lattice, generating point defects, vacancies, and deep-level traps that degrade carrier transport and junction characteristics[6-10]. Prior studies have shown that neutron irradiation can cause significant performance degradation in β-$Ga_2O_3$ devices, particularly through carrier compensation and increased trap-assisted recombination[9-14]. However, these studies also suggest an inherent degree of resilience in β-$Ga_2O_3$ under extreme radiation exposure, which opens the door to new engineering strategies. While much of the existing research has focused on degradation mechanisms, there is growing interest in exploring how controlled irradiation could be strategically leveraged to tune or even enhance device characteristics, highlighting a promising and largely untapped area of study. Thermal annealing, in particular, has been shown to partially restore electrical performance of β-$Ga_2O_3$ by passivating radiation-induced defects, recovering carrier density, and healing lattice disruptions[15-17]. Recent Deep-Level Transient Spectroscopy (DLTS) studies have shown that radiation-induced defect states in β-$Ga_2O_3$ can begin to anneal at temperatures as low as ~200 °C and continue to evolve up to 400 °C, indicating that thermal annealing processes can influence defect behavior in irradiated β-$Ga_2O_3$[18]. While conventional annealing can lead to trade-offs, such as contact degradation and increased leakage[19-21], electro-thermal annealing, where thermal energy is introduced during active device operation, can potentially be more targeted and dynamic approach. Notably, performance recovery and enhancement effects have been observed in β-$Ga_2O_3$ devices subjected to high-dose gamma irradiation, serving effectively as radiation annealing, or proton irradiation followed by thermal annealing[5, 18, 22-25], yet systematic investigations involving fast neutron irradiation with electro-thermal annealing remain largely unexplored.

In this study, we combine fast neutron irradiation with controlled electro-thermal annealing to assess, restore and significantly enhance the voltage-blocking capabilities of β-$Ga_2O_3$ SBDs, achieved without employing any additional field-management or edge-termination techniques. Devices were irradiated with 1 MeV neutrons to a total fluence of $1\times10^{15}$ n/cm² and subsequently subjected to repeated electro-thermal testing cycles in which the device response was measured at temperatures from 25 °C to 250 °C in fixed 50 °C increments. Following irradiation, the diodes initially exhibited significant degradation in forward conduction and on-resistance; however, substantial recovery was achieved through electro-thermal biasing. The breakdown voltage improved significantly, with capacitance-voltage (C-V) and current-voltage (J-V) analyses confirming carrier compensation and trap annealing as the underlying mechanisms. These results not only advance our understanding of defect dynamics in neutron-irradiated β-$Ga_2O_3$ but also demonstrate a viable pathway for engineering radiation-hardened β-$Ga_2O_3$ power devices through neutron-assisted defect modulation and electro-thermal recovery.

The SBDs was fabricated on halide vapor phase epitaxy (HVPE) grown ~10.8 μm thick Si-doped $n^-$ β-$Ga_2O_3$ drift layer (~$3\times10^{16}$ cm$^{-3}$) on a Sn doped $n^+$ (~$5.4\times10^{18}$ cm$^{-3}$) (001) β-$Ga_2O_3$ native substrate. A schematic of the diode structure is shown in Figure 1(a). The fabrication process started with $BCl_3$-based reactive-ion etching to remove 1μm of $Ga_2O_3$ from the backside. This was followed by the deposition of a Ti/Au ohmic metal stack using electron beam evaporation, which was then annealed at 470˚C for 1 minute through rapid thermal annealing (RTA). Lastly, the top Ni/Au Schottky contacts were patterned using electron beam lithography. Other than the standard rapid thermal annealing applied to the ohmic contact, no thermal annealing was performed after completing the full device fabrication process before irradiation. Post-fabrication testing included J-V measurements conducted with an HP 4155B semiconductor parameter analyzer, along with reverse breakdown characterization and C-V measurements using an Agilent 4294A precision impedance analyzer. Neutron irradiation was performed at the University of Massachusetts Lowell Radiation Laboratory, utilizing a 1 MeV neutron source with a flux of $8\times10^{10}$ n/cm$^2$-s to achieve a total fluence of $1\times10^{15}$ n/cm$^2$. Post-irradiation, electro-thermal testing was performed to evaluate the combined effect of elevated temperature and applied bias on device recovery. The testing was carried out in four electro-thermal cycles, labeled as electro-thermal testing 1, 2, 3, and 4 in the figures. In each cycle, the device temperature was increased in fixed increments of 50 °C, labeled as electro-thermal testing 1, 2, 3, and 4 in the figures and the text. In each cycle, the device temperature was increased in fixed increments of 50 °C, and current-voltage measurements were recorded at 25 °C, 50 °C, 100 °C, 150 °C, 200 °C, and 250 °C. After the measurement at 250 °C, the device was allowed to return to room temperature by natural cooling before concluding the cycle. Electro-thermal testing 1 corresponds to the first complete temperature sequence following neutron irradiation, and testing 2, 3, and 4 correspond to repetitions of the same procedure. In all electro-thermal cycles, the device remained at each temperature only for the duration required to complete the I-V measurement, and the total thermal exposure time was identical from cycle to cycle.

Figures 1(b)-1(d) show the room-temperature J-V characteristics of the devices under various conditions, including the pre-irradiation state, the post-irradiation state, and after successive electro-thermal testing cycles. After irradiation, all devices initially exhibited a drastic reduction in the on-current, indicating substantial degradation in their electrical performance. However, after the first electro-thermal testing, devices showed a noticeable recovery, with a significant increase in the on-current. After the second electro-thermal testing, the on-current showed further improvement and then stabilized during the third and subsequent electro-thermal cycles, indicating that the recovery process had reached saturation once the majority of thermally responsive defects had been annealed. A similar trend is observed in the Schottky barrier height, SBH ($\Phi_B$), extracted from the forward J-V characteristics considering thermionic emission as the dominant transport mechanism at room temperature[23, 26-29]: The SBH increases from 1.06 eV in the pre-irradiated device to 1.13 eV after the first electro-thermal cycle, and remains near 1.12 eV in the subsequent cycles, consistent with prior observations of irradiation-induced increases in the effective barrier height of β-$Ga_2O_3$ Schottky diodes[5, 11, 29]. This trend is also consistent with the SBH extracted

independently from C-V analysis and temperature-dependent J-V measurements, as discussed in later sections. This stability in SBH after the initial recovery stage indicates that the junction properties reach equilibrium once the temperature-activated defect relaxation has occurred. Similarly, the specific on-resistance $R_{on\text{-}sp}$ demonstrated a sharp increase immediately after irradiation, reflecting radiation-induced damage. However, as electro-thermal testing was performed, $R_{on\text{-}sp}$ progressively decreased and trended toward pre-irradiation levels. This gradual restoration of $R_{\text{on-sp}}$, coupled with the stabilization of the on-current, indicates that the electro-thermal testing effectively mitigates radiation-induced defects and restores the device performance. These degradation and recovery trends were consistently observed across all devices with different active areas as shown in Figures 1(b)-(d), confirming the reproducibility of the behavior and supporting the reliability of the overall observations.

To investigate the effects of electro-thermal testing on device capacitance, built-in potential ($V_{bi}$), SBH ($\Phi_B$) and net carrier concentration ($N_d^+$ - $N_a^-$), C-V measurements were performed after each stage of testing. A significant reduction in capacitance was observed following neutron irradiation as shown in Figure 2(a), with the C-V curve becoming nearly flat, indicating severe radiation-induced damage to the device. After the first electro-thermal testing, the capacitance increased noticeably, trending toward pre-irradiation levels. Subsequent electro-thermal testing further enhanced the capacitance, which eventually stabilized but remained slightly reduced compared to the pre-irradiation state. The $V_{bi}$, $\Phi_B$ and $N_d^+$-$N_a^-$ are determined using the following formulas using the relative permittivity of β-$Ga_2O_3$, $\varepsilon_r$=10, density of states in the conduction band, $N_C$ = 5.2×$10^{18}$ $cm^{-3}$, where $A$ representing the device area[23, 26-29].

$$N_d^+ - N_a^- = \frac{2}{q\varepsilon_r\varepsilon_0 A^2\left(\frac{d\frac{1}{C^2}}{dV}\right)} \quad (1)$$

$$\frac{A^2}{C^2} = \left[\frac{2}{q\varepsilon_r\varepsilon_0\left(N_d^+ - N_a^-\right)}\right]\left(V_{bi} - \frac{kT}{q} - V\right) \quad (2)$$

$$\varphi_B = V_{bi} + \frac{KT}{q} ln\left[\frac{N_C}{N_d^{+} - N_a^{-}}\right] \quad (3)$$

As shown in Figure 2(b), $V_{bi}$ extracted from the $1/C^2$-$V$ characteristics decreases slightly after the first post-irradiation electro-thermal cycle from 0.988 V to 0.968 V, and then stabilizes near 0.978 V during subsequent testing. Using these $V_{bi}$ values, the SBH (q$\Phi_B$) was calculated, yielding 1.13 eV before irradiation, increasing to 1.15 eV after the first electro-thermal cycle, and remaining near 1.14 eV in subsequent cycles, which closely tracks the SBH obtained from forward J-V analysis. The net carrier concentration as shown in Figure 2(c), derived from the C-V profiles, showed a significant decrease from $3.2\times10^{16}$ cm$^{-3}$ to $5.5\times10^{15}$ cm$^{-3}$ after the first electro-thermal testing, with a carrier removal rate of 26.5 cm$^{-1}$. This reduction can be attributed to the annealing of radiation-induced traps, which initially acted as compensating centers, reducing the free carrier density. Subsequent electro-thermal testing resulted in an increase in carrier concentration $9.9\times10^{15}$ cm$^{-3}$, reflecting a carrier removal rate of 22.1 cm$^{-1}$, and eventually stabilized after continued annealing. This partial recovery is likely due to further healing of radiation-induced defects and the reactivation of donor sites within the material. Previous neutron-irradiation studies on β-$Ga_2O_3$ employing deep-level transient spectroscopy (DLTS) and deep-level optical spectroscopy (DLOS) report strong carrier compensation with carrier-removal rates ranging from ~19 cm$^{-1}$ to over 50 cm$^{-1}$ for 1 MeV equivalent neutron fluences, depending on irradiation fluence, growth method, doping, dopants, and defect spectrum[9, 10]. These works consistently identify a set of irradiation-sensitive deep levels that dominate the compensation process. In plasma-assisted molecular beam epitaxy grown Ge-doped β-$Ga_2O_3$ films, neutron irradiation introduces or enhances defects near $E_C$ - 0.78 eV, $E_C$ -1.22 eV, and $E_C$ - 2.00 eV within the bandgap, which have been attributed to gallium vacancies, vacancy-interstitial complexes, hydrogenated vacancy complexes, or Ga-O antisites[9]. In unintentionally doped edge-defined film-fed grown β-$Ga_2O_3$ substrates, neutron-induced compensation is primarily associated with newly introduced $E_C$ - 1.29 eV traps together with an enhanced $E_C$ - 2.00 eV level, both identified as the dominant compensating defects[10]. The substantial reduction in net carrier concentration observed in this work after irradiation is therefore consistent with these previously established compensating defect signatures, confirming that neutron-induced deep levels of this type are responsible for the strong carrier removal seen in β-$Ga_2O_3$ under displacement damage.

The impact of electro-thermal testing and temperature-dependent recovery of device performance after irradiation is shown in Figure 3. The pre-radiation J-V characteristics shows high on-currents (~890 A/cm²) and low specific on-resistance (~3.14 mΩ·cm²). After neutron irradiation, a dramatic reduction in the on-current to ~$4.65\times10^{-6}$ A/cm² and an increase in $R_{on,sp}$ to ~$7.8\times10^{8}$ mΩ·cm² were observed at 25˚C as shown in Figures 3(a-b), indicating significant radiation-induced damage. These changes are attributed to the formation of deep-level traps and compensating defects, which reduce carrier mobility and concentration, resulting in degraded device performance. During the first electro-thermal testing, as the temperature increased from 25˚C to 250˚C, significant improvements in device performance were observed. The on-current increased steadily, reaching from $4.65\times10^{-6}$ A/cm² (post-radiation) to ~3.7 A/cm² at 250˚C, while $R_{on,sp}$ decreased from ~$7.8\times10^{8}$ mΩ·cm² (post-radiation) to ~1562 mΩ·cm². This progressive recovery with temperature suggests that increasing thermal energy enables the sequential activation and annealing of both shallow and deep-level radiation-induced traps, leading to enhanced carrier transport and a marked improvement in device performance by 250 °C, as reflected in Figures 3(a) and (b). The second electro-thermal testing as shown in Figures 3(c) and (d) continued the recovery process but with diminishing returns, indicating that most recoverable defects had already been mitigated during the first cycle. As the temperature increased from 25˚C to 250˚C during second electro-thermal testing, the on-current improved further, reaching from ~20 A/cm² to ~83 A/cm², and $R_{on,sp}$ decreased to ~47 mΩ·cm² from ~770 mΩ·cm². During the third electro-thermal testing, as shown in Figures 3(e) and (f), further improvements in device performance were observed. The on-current increased to ~140 A/cm², while the $R_{on,sp}$ decreased to approximately 26 mΩ·cm². However, as the temperature increased from room temperature to 250°C, the on-current slightly reduced from 140 to 117 A/cm², and $R_{on,sp}$ increased from 26 to 33 mΩ·cm². This behavior is more reflective of pre-radiation trends, differing from the earlier electro-thermal cycles 1 and 2, where device performance improved consistently with increasing temperature. This finding suggests that the devices underwent substantial recovery, and their characteristics began to exhibit trends similar to pre-radiation conditions. To confirm the stability of this recovery, a fourth electro-thermal testing cycle was conducted, as shown in Figures 3(g) and (h). During this cycle, similar trends of decreasing on-current and increasing $R_{on,sp}$ with rising temperature were observed. The on-current and $R_{on,sp}$ showed almost identical values to those recorded during the third testing, indicating that most thermally recoverable traps had been annealed during the earlier cycles. The observed reduction in current density with increasing temperature is consistent with reduced electron mobilities in the current-limiting series resistance of the device at higher temperatures[30, 31]. The negligible changes during the fourth testing highlight a saturation point in the recovery process, where further improvements are limited. Across all four electro-thermal testing cycles, the on-current and resistance exhibited strong temperature-dependent behavior, with significant improvements during the first two cycles and stabilization in subsequent cycles. The on-current increased from approximately $4.65\times10^{-6}$ A/cm² post-irradiation to ~180 A/cm² after the fourth cycle, while $R_{on,sp}$ decreased from approximately $7.8\times10^{8}$ mΩ·cm² to approximately 21 mΩ·cm², indicating the effectiveness of electro-thermal testing in mitigating radiation-induced damage. The fact that the recovered on-resistance after post-irradiation electro-thermal annealing remains slightly higher than the pre-irradiation value (~3.14 mΩ·cm²) is consistent with the reduced net carrier concentration extracted from the C-V measurements (Fig. 2(c)), since the drift-layer resistance scales inversely with the net carrier concentration[32].

The temperature-dependent behavior of SBH ($\Phi_B$), ideality factor ($\eta$), leakage current, and rectification ratio ($I_{on}/I_{off}$) across different electro-thermal testing cycles is presented in Figures 4(a)-4(d). The $\eta$ and SBH can be extracted using the standard thermionic-emission (TE) formalism[23, 26-29], applied to the measured temperature-dependent J-V characteristics,

$$J = J_s\left[exp\left(\frac{qV}{\eta k_0 T}\right) - 1\right] \quad (4)$$

$$J_s = A^* T^2\ exp\left(-\frac{q\varphi_B^{JV}}{k_0 T}\right) \quad (5)$$

$$A^* = \frac{4\pi q m_n^* k_0^2}{h^3} \quad (6)$$

where $q$ is the electric charge, $k_0$ is the Boltzmann constant, and $\eta$ is the ideality factor, $J_s$ is the reverse saturation current density, $\varphi_B^{JV}$ is the apparent SBH obtained from the TE model, and $A^*$ is Richardson's constant, which is calculated to be 41.04 $Acm^{-2}K^{-2}$ [27, 30]. Since the accuracy of $\varphi_B^{JV}$ extracted from Eq. (5) depends strongly on $\eta$, especially when $\eta > 1$, a correction was applied, as described by Wagner et al.[33], for the SBH calculation[27, 28, 33-35] using following equation[33]:

$$\varphi_B = \left(\varphi_B^{JV} - \frac{\eta - 1}{\eta}\frac{kT}{q} ln\frac{N_C}{N_D}\right)\eta \quad (7)$$

where $N_C$ is the effective conduction-band density of states and $N_D$ is the donor concentration obtained from C-V measurements. This correction provides a more reliable estimate of the true SBH in cases where interface recombination or defect-assisted transport increases the apparent ideality factor, as is typical for irradiated β-$Ga_2O_3$ diodes.

Figure 4(a) shows the temperature-dependent evolution of the ideality factor for the pre-irradiated and post-irradiated device across the different electro-thermal testing cycles. Pre-radiated devices showed excellent ideality factors ranging between 1.01-1.05 across all investigated temperatures. After radiation, the ideality factor at room temperature increases significantly to 2.00, indicating the presence of radiation-induced traps and recombination centers at the metal-semiconductor interface. During the first electro-thermal cycle, the ideality factor decreases steadily with temperature, dropping from 2.00 at 25 °C to 1.21 at 250 °C, as thermal activation facilitates the annealing of radiation-induced defects. This reduction reflects the effective recovery of the interface properties during the initial cycle. In the second and third electro-thermal testing cycles, the ideality factor shows minimal variation across the temperature range, stabilizing close to pre-radiation levels. The extracted SBH as shown in Figure 4(b) shows a slight decrease with increasing temperature for all electro-thermal testing cycles. This negative temperature dependence is consistent with previous studies on β-$Ga_2O_3$ Schottky diodes [26, 31, 36]. After irradiation, the SBH exhibits a slight increase across all investigated temperatures compared to the pre-radiation values, consistent with previous reports[5, 11, 29]. Figure 4(c) shows the temperature-dependent behavior of on-resistance across different electro-thermal testing cycles. During the first cycle, $R_{on,sp}$ decreases significantly at higher temperatures, reflecting the annealing of radiation-induced traps and the partial recovery of carrier transport. The second cycle continues this trend, with further reductions in $R_{on,sp}$ as temperature increases. In the third cycle, however, $R_{on,sp}$ begins to increase with temperature, mirroring the behavior observed in the pre-radiation condition. This transition indicates that the device has undergone substantial recovery, with its transport characteristics again dominated by the temperature-dependent mobility trends of β-$Ga_2O_3$ rather than radiation-induced defects. The leakage current, as shown in Figure 4(d), decreases after neutron irradiation, likely due to a reduction in the surface electric field caused by the lower net carrier concentration resulting from the compensation by radiation-induced traps[11]. Additionally, the slight increase in SBH observed post-irradiation (Figure 4(b)) could contribute to the reduced leakage current by suppressing thermionic emission[37], as fewer carriers have sufficient energy to overcome the higher barrier. While the leakage current remains consistently lower than pre-radiation levels, which is in good agreement with the behaviors observed in proton and neutron irradiated devices[11, 13, 38], it increases across all conditions at higher temperatures due to enhanced carrier activation and thermionic emission over the Schottky barrier. Electro-thermal testing shows minimal impact on leakage current across different cycles, with only slight variations observed. The rectification ratio ($I_{on}/I_{off}$), as shown in Figure 4(d), decreases as the temperature increases for pre-radiation condition. During the first electro-thermal testing, a significant reduction in the rectification ratio was observed at room temperature compared to pre-radiation conditions, reflecting the degradation caused by neutron irradiation. However, as the temperature increased during the first testing cycle, the rectification ratio improved significantly. In the second and third electro-thermal testing cycles, the rectification ratio showed substantial recovery at room temperature, approaching pre-radiation levels.

The reverse breakdown characteristics ($V_{br}$) of the devices following neutron irradiation and electro-thermal testing was also investigated, as shown in Figure 5. The breakdown voltage shows a significant enhancement after neutron irradiation and the first electro-thermal testing. Specifically, the breakdown voltage increases dramatically into the kilovolt class, reaching 1.28 kV from ~300 V pre-radiation, corresponding to an increase of over 325%. This substantial enhancement is attributed to a drastic reduction in net carrier concentration by more than 82% as shown in Figure 2(c), primarily caused by the compensation of carriers due to radiation-induced traps. In subsequent electro-thermal testing cycles, the breakdown voltage decreases slightly, reaching up to 940 V after the third cycle. Further electro-thermal cycles produced nearly identical breakdown characteristics, reflecting the stabilization of the net carrier concentration at its partially recovered value and the minimal change introduced by additional thermal cycling. This behavior is consistent with the partially compensated state of the drift region after electro-thermal treatment, where the net carrier concentration remains well below the pre-irradiation value. Nevertheless, the breakdown voltage remains significantly higher than the pre-radiation level and approaches the kilovolt class, even in the absence of any field management techniques.

In summary, this study demonstrated the impact of fast neutron irradiation and electro-thermal annealing on the electrical performance of β-$Ga_2O_3$ SBDs. Neutron irradiation initially caused substantial degradation, including a significant reduction in on-current and an increase in specific on-resistance. A drastic decrease in net carrier concentration is also observed due to the formation of radiation-induced traps. Electro-thermal testing resulted in significant recovery of device performance. Key observations include a remarkable enhancement in breakdown voltage, which increased from 300V to 1.28 kV after the first electro-thermal cycle, driven primarily by an > 82% reduction in net carrier concentration due to trap compensation. Subsequent electro-thermal cycles showed diminishing improvements, with the breakdown voltage stabilizing up to 940 V, consistent with the increase of net carrier concentrations in subsequent electro-thermal testing. Similarly, the on-current and specific on-resistance exhibited substantial recovery during the first two testing cycles, followed by stabilization in later cycles, indicating that the majority of thermally responsive defects had already been annealed. Overall, these findings demonstrate that neutron irradiation combined with controlled electro-thermal annealing can serve as an effective approach for mitigating radiation-induced damage and significantly enhancing the voltage-blocking capability of β-$Ga_2O_3$ power devices.

**Acknowledgments**

The authors acknowledge the funding support from AFOSR under award FA9550-18-1-0479 (Program Manager: Ali Sayir), NSF awards ECCS 2019749, 2231026, 1919798, 2501623 and 2532898, ARPA-E award DE-AR0001879 and Coherent II-VI Foundation Block Gift Program.

**Conflict of Interest**

The authors have no conflicts to disclose.

**Data Availability**

The data that support the findings of this study are available from the corresponding author upon reasonable request.

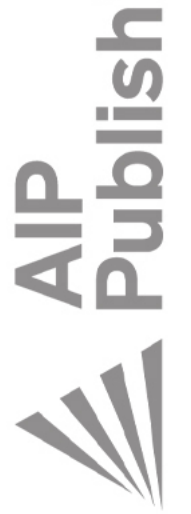

**Figure 1**

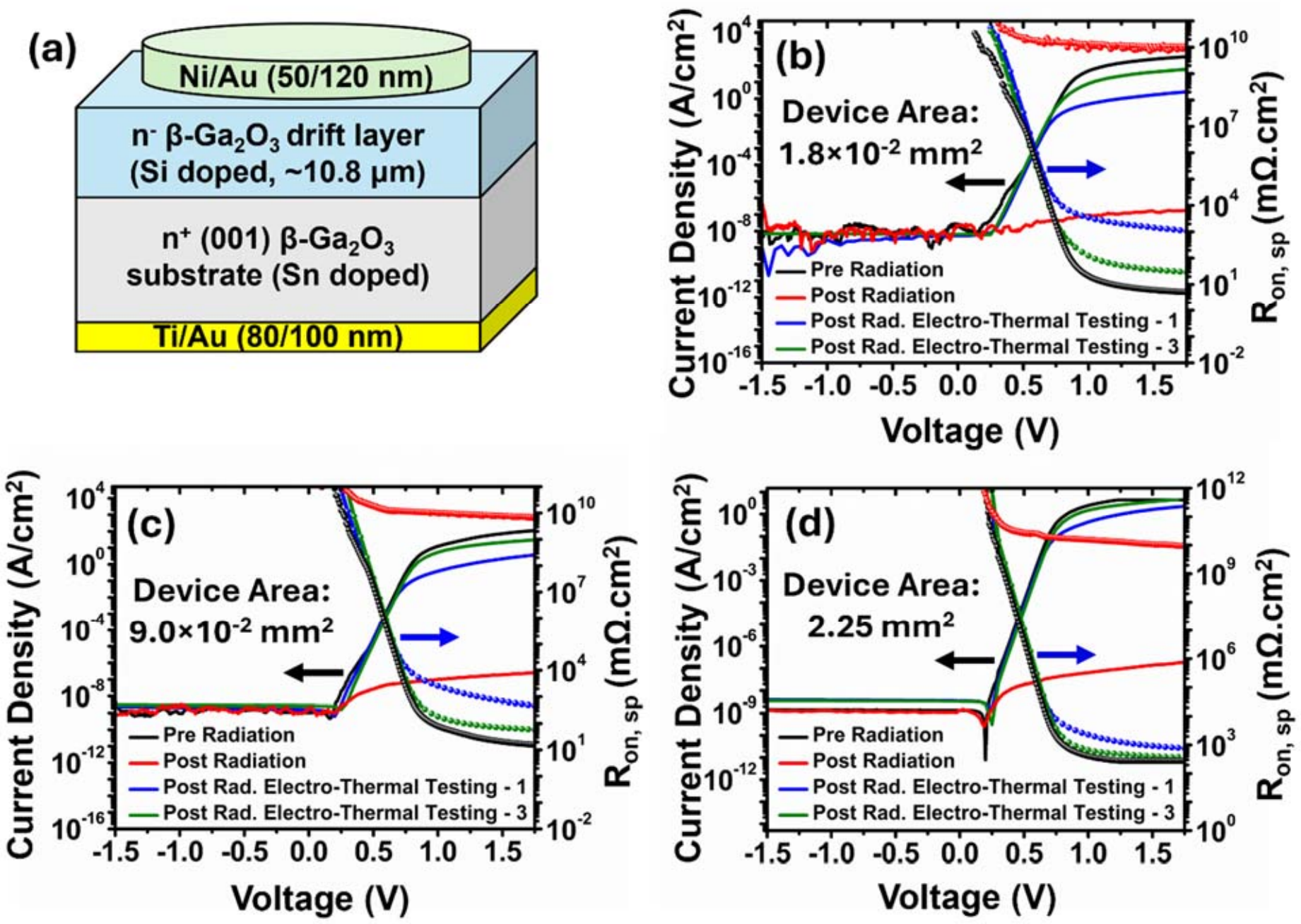

**Figure 1.** (a) Schematic cross-section of the β-$Ga_2O_3$ Schottky barrier diode (b-d) room temperature current-voltage (J-V) characteristics and specific-on-resistances of the devices with varying areas, shown for pre-radiation conditions and after electro-thermal testing cycles following neutron irradiation.

**Figure 2**

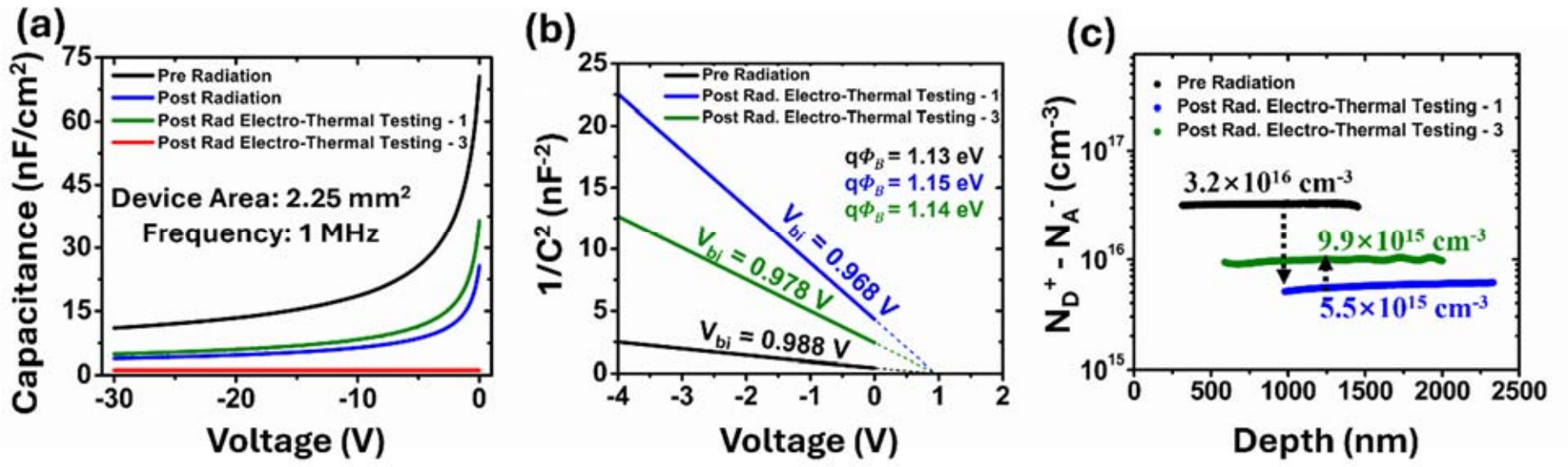

**Figure 2.** (a) Comparison of capacitance-voltage (C-V) characteristics for pre- and post- radiation conditions after different electro-thermal testing cycles, (b) $1/C^2$-V plot used to determine the built-in potential ($V_{bi}$), and (c) net carrier concentration profile as a function of depth, derived from the C-V curves.

**Figure 3**

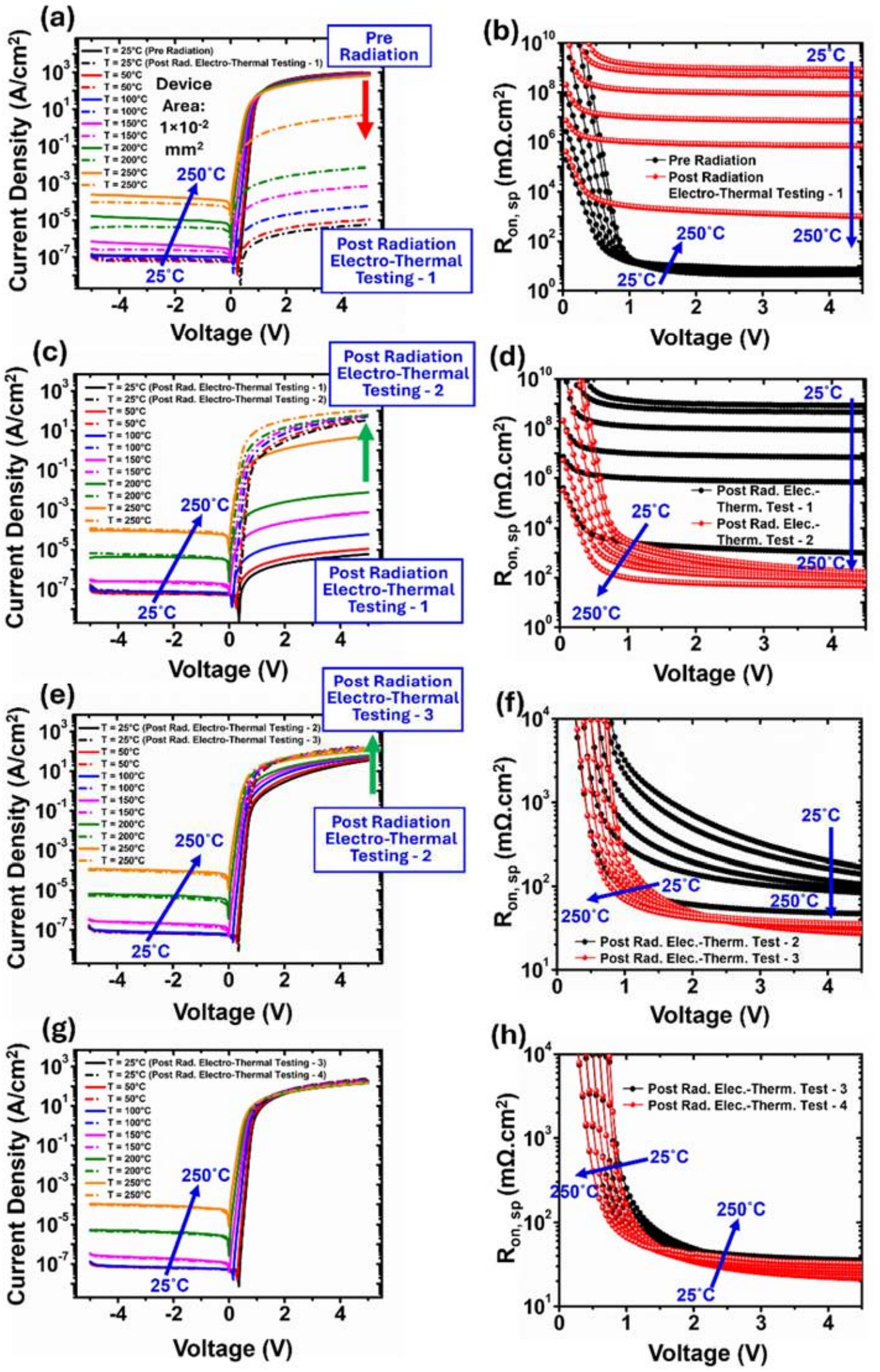

**Figure 3.** (a, c, e, g) Temperature dependent J-V characteristics and (b, d, f, h) specific-on-resistance vs. voltage characteristics of the diodes for pre-radiation conditions and during different electro-thermal testing cycles following neutron irradiation.

**Figure 4**

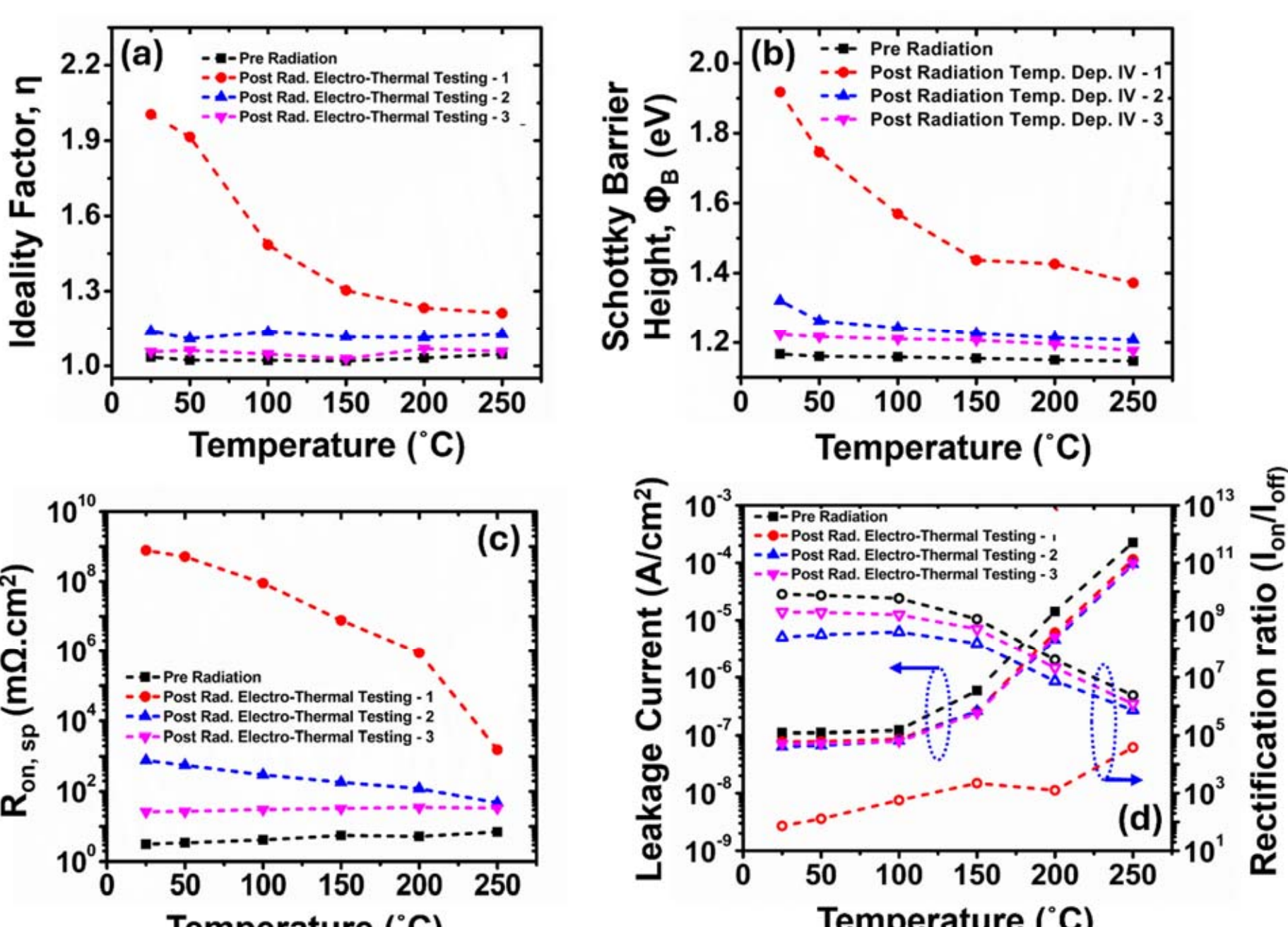

**Figure 4.** (a) ideality factor, (b) Schottky barrier height, (c) specific on-resistance, (d) reverse leakage current and rectification ratio ($I_{on}/I_{off}$) of the diode as a function of temperature, measured before irradiation and during each electro-thermal testing cycle following neutron irradiation.

**Figure 5**

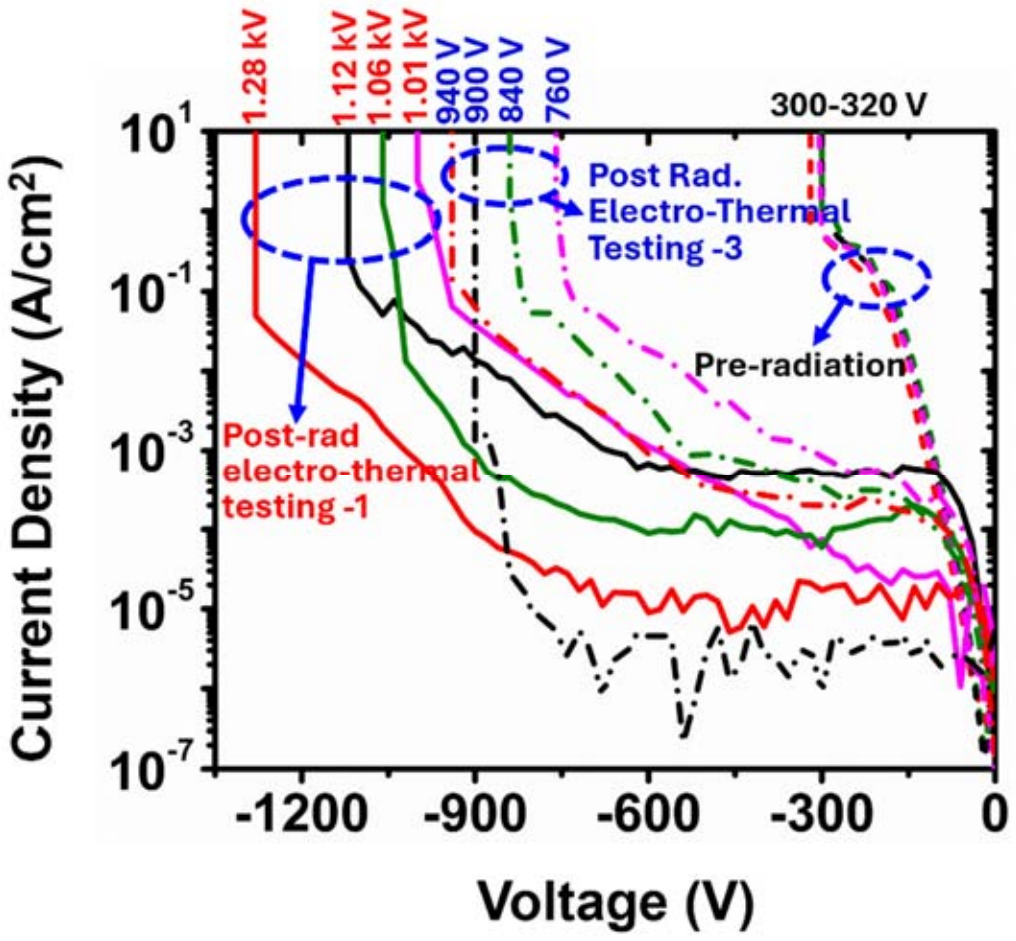

**Figure 5.** Reverse J-V characteristics of the devices showing breakdown voltages: before irradiation and after electro-thermal testing cycles following neutron irradiation.